\documentclass[preprint,showpacs,preprintnumbers,amsmath,amssymb,superscriptaddress]{revtex4}

\usepackage{graphicx,subfigure}
\usepackage{dcolumn}
\usepackage{bm}
\newcommand{\la}{\left\langle}
\newcommand{\ra}{\right\rangle}

\begin{document}
\title{Phase separation instabilities and pairing modulations in $Bi_2Sr_2CaCu_2O_{8+\delta}$}

\author{Kun Fang}
\affiliation{Department of Physics, University of Connecticut, Storrs, CT 06269, USA}
\author{G. W. Fernando}
\affiliation{Department of Physics, University of Connecticut, Storrs, CT 06269, USA}
\author{A. Balatsky}
\affiliation{Theoretical Division and Center for Integrated Nanotechnologies, Los Alamos National Laboratory, Los Alamos, NM 87545, USA}
\author{A. N. Kocharian}
\affiliation{Department of Physics, California State University, Los Angeles, CA 90032, USA}
\author{Kalum Palandage}
\affiliation{Department of Physics, Trinity College, Hartford, CT 06106, USA}

\begin{abstract}
There is  growing evidence that the unconventional spatial inhomogeneities in the doped high-Tc superconductors are accompanied by the pairing of electrons, subsequent quantum phase transitions (QPTs), and condensation in coherent states. We show that these superconducting states can be obtained from  phase separation instabilities near the quantum critical points. We examine electron coherent and incoherent pairing instabilities using our results on exact diagonalization in pyramidal and octahedron Hubbard-like clusters under variation of chemical potential (or doping), interaction strength, temperature and magnetic field. We also evaluate the behavior of the energy gap in the vicinity of its sign change as a function of out-of-plane position of the apical oxygen atom, due to vibration of apical atom and variation of inter-site coupling. These results provide a simple microscopic explanation of (correlation induced) supermodulation of the coherent pairing gap observed recently in the scanning tunneling microscopy experiments at atomic scale in $Bi_2Sr_2CaCu_2O_{8+\delta}$. The existence of possible modulation of local charge density distribution in these materials is also discussed.

\end{abstract}



\maketitle

\section{Introduction}~\label{intro}

Since the discovery of the high-temperature (high-$T_c$) superconductors in 1986, there have been many advances in the field of superconductivity due to intense worldwide research efforts. Several sets of ceramic cuprate materials have been synthesized and the transition temperature $T_c$ in some of these has exceeded $140K$. They all share a common quasi-two-dimensional structure and the superconductivity is believed to originate from the electron correlations in the $CuO_2$ planes. In spite of intensive study of these compounds during the past decade, there is no consensus on a possible mechanism for superconductivity. At the same time, researchers have focused on out-of-plane effects as an explanation in the cuprate superconductors, because all the cuprate families share the same $CuO_2$ plane, but the maximum $T_C$ varies dramatically from one cuprate family to another~\cite{Eisaki}. This variation is unlikely to be caused by the electronic interactions in the $CuO_2$ plane, but it is most probably due to different atomic structures between the planes and their indirect effect on the electronic structure in the $CuO_2$ plane.

Recently, it has been argued that the non-planar apical oxygen atoms in the $CuO_5$ pyramid or $CuO_6$ octahedron play an important role in the out-of-plane atomic-scale lattice effects in high $Tc$ superconductors (HTSCs)~\cite{Eisaki,Zhou,Merz,Ohta}. More systematic experimental studies have been done on $Bi_2Sr_2CaCu_2O_{8+\delta}$~\cite{Slezak, Balatsky, Kan} ($Bi$-$2212$). Because of the natural cleavage plane between two $BiO$ layers, the above system has been ideal for probing the planar electronic structure. This material also has a complicated superstructure modulation (supermodulation) which is known to have a strong effect on the position of the apical oxygens but almost no effect on the $CuO_2$ plane~\cite{Kan}. Slezak et al. measured the local superconducting gap using STM with atomic precision and found that the energy gap also has a modulation with the same periodicity as the structural supermodulation~\cite{Slezak}.

Superconductivity in cuprate materials is often discussed in the context of the quantum critical point that controls the optimal doping and maximum in the transition temperature~\cite{Sachdev,Sondhi}. Most measurements of magnetic correlation length near putative quantum critical point (QCP) reveal that it remains finite. Scanning probes ubiquitously reveal electronic nanoscale inhomogeneity (See Ref.~\cite{Slezak} and references therein). All these findings cause us to question the assumption of a smooth evolution to the quantum critical point. A possible alternative explanation of these findings is that the electronic matter develops a spontaneously inhomogeneous modulations (stripes, checkerboards, etc.) near QCP~\cite{She}. This propensity to spontaneous charge inhomogeneity can significantly benefit from the electronic modulations induced by the apical oxygen positions. From this perspective it is natural to expect that apical oxygen position modulations provide a perturbation that leads to in-plane charge and spin modulation and therefore would lead to a spatially modulated superconducting state with finite correlation lengths both for charge and spin excitations. In addition, scanning tunneling microscopy (STM) measurements in underdoped $Bi$-$2212$ provide information about two energy scales or gaps, coherent vs. incoherent pairings, metal-insulator transition, etc. driven by temperature and electron concentration~\cite{Sebastiana,Yazdani,Hudson,Tanaka,Lee}.

For most HTSCs, the changes in physical properties are obtained by doping carriers through changes of composition. Apart from the usual doping cases, some HTSCs with Bi, such as $Bi$-$2212$, can experience self-doping, where the ions in non-conducting layers can act as a charge reservoir, which provides free carriers to the $CuO_2$ plane~\cite{Khomskii}. Here we investigate possible supermodulation of a charge redistribution (self-doping) between the charge reservoir and the $CuO_2$ plane at fixed stoichiometry. Inspired by the observation of gap modulation, several theoretical proposals have been put forward by introducing variations of coupling constants on a phenomenological level, or applying the conventional BCS theory~\cite{Nunner, Anderson, Yang, Graser, Machida}. Here we report the results of a microscopic study of pairing gap modulation for repulsive electrons within individual unit cells in real space for various spatial displacements of the apical atoms on superconducting pairing.

In this paper we discuss the physical nature of the observed local pairing gap modulation and possible charge redistribution in connection with the structural changes, the change of chemical potential and self-doping effect. We argue that the superconducting pairing correlations are strongly driven by pairing modulations while charge density and the chemical potential in the supermodulation phase can change slightly with the periodicity of structural changes. The paper is organized as follows: Following the introduction to the various aspects of pairing modulations, we present the microscopic model and methodology of exact cluster calculations of pyramidal structures in section~\ref{Method}. In section~\ref{result}, we show the results of charge gap variations and pairing modulation in $Bi_2Sr_2CaCu_2O_{8+\delta}$. The concluding summary constitutes Section~\ref{summary}. In Appendixes, we present the details of calculations related to various aspects
of modulation properties.

\section{Model}~\label{model}

Slezak et al. showed a direct link between the spatial modulation of the out-of-plane atomic structure within an individual unit cell and the variation of the local gap maximum~\cite{Slezak}. The non-planar atoms, especially the apical oxygen atoms, are playing an indirect but important role on modulating the local electronic structure of the $CuO_2$ plane. Merz et al.~\cite{Merz} concluded from their X-ray absorption spectroscopy that increasing $O$ dopant can induce holes both on oxygens in $CuO_2$ plane and the apical oxygen atom. This result suggests that both the planar and apical oxygens can be electronically connected and might be controlling the local electronic state by exchanging holes between $CuO_2$ plane and other planes. This could be the origin of interlayer interactions and local gap variations. Therefore, in order to simulate the spatial gap modulation, we need a model which can treat the local electronic state exactly and simultaneously include an out-of-plane interaction that can exchange electrons/holes with the basal atoms. This requirement suggests the use of a quantum cluster treatment to deal with local correlations and an extra out-of-plane site to control the electron/hole concentration on the $CuO_2$ plane.

\begin{figure}
\includegraphics*[width=20pc]{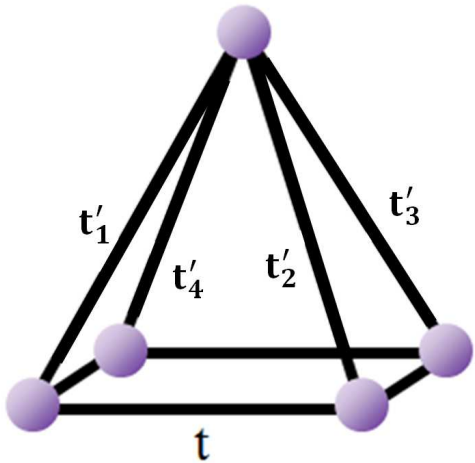}
\caption{A schematic picture of the pyramid cluster discussed in this work: the hopping parameters on the basal plane are denoted by $t$ and the hopping parameters between the top site and basal sites are $t'_1,t'_2,t'_3,t'_4$. Here, all the energy parameters are measured in units of $t$.}
\label{fig:fig1}
\end{figure}

We begin with the canonical single band Hubbard model on a pyramid cluster (as shown in Fig.~\ref{fig:fig1}).
\begin{eqnarray}
H=-t\sum\limits_{\la i,j\ra\sigma}(c^+_{i\sigma}c_{j\sigma}+H.c.)-\sum\limits_{i\sigma}t'_i(a^+_\sigma c_{i\sigma}+H.c.)+U \sum\limits_{i} n_{i\uparrow}n_{i\downarrow}
\label{eqn:h}
\end{eqnarray}
Here, $c_{i\sigma}$ ($c^{+}_{i\sigma}$) is the electron destruction (creation) operator at the basal sites with spin $\sigma$ (or magnetic sublevel), while $a_\sigma$ ($a^+_\sigma$) is the same operator at the apical site. The first term indicates the hopping between the nearest basal sites $i$ and $j$ ($\la i,j\ra$ denotes the nearest neighbor sites) with the same hopping parameter $t$.  The second term describes hopping between the apical site and the basal site $i$ with a hopping amplitude $t'_i$. In addition, $U>0$ is the on-site Coulomb interaction. In practice, the hopping within the plane should be much bigger than that of the out-of-plane atom, {\it i.e.} $t'_i\ll t$. The energies are measured with respect to $t>0$, unless otherwise stated. The cluster can be solved exactly for all the eigenvalues and eigenvectors~\cite{PRB_06}. The apical site can exchange holes with all the basal sites, so displacements of the top site within the cluster will affect the planar electronic structures.

In the absence of electron-hole symmetry,
the sign of the hopping amplitude between the nearest neighbor sites leads to distinct changes of the electronic structure~\cite{PRB_08}. In nonbipartite structures, such as the pyramid or tetrahedron, the results strongly depend on the sign of $t$ and $t'_i$. Here the pyramidal non-bipartite geometry is defined by choosing $t>0$ as the hopping connecting square (plaquettes) sites in the basal plane, and we consider various coupling parameters $0<t'_i\ll t$ as hoppings between the apical site and the planar site $i$ in the square base (Fig.~\ref{fig:fig1}).

\section{Methodology}~\label{Method}

An essential element for understanding our results on pairing modulation is the formation of coherent pairing state of $n$ electrons with a negative charge gap and positive spin gap obtained earlier in Refs.~\cite{PRB_06,PRB_08,PLA_09,PRB_09,PLA_12,JMMM_12,Springer} and references therein. For completeness here we first briefly summarize the key results of exact diagonalization in small clusters at weak and moderate $U$ values. The square lattices, generated recently on optimized (finite) Betts unit cells, provide strong evidence for the phase separation instabilities found in generic $2\times 2$, $2 \times 4$ and other clusters~\cite{PLA_12,JMMM_12}.

\subsection{Charge and spin gaps}

Numerical (exact) calculations of the energy levels $E_n$ for the relevant electron number $n$ and a given spin $S$ are used to define  charge and spin gaps. The charge excitation gap $\Delta^c$ at finite temperature can be written as an energy difference of canonical energies, $\Delta^c(n,T)=\mu_+-\mu_-$, where $\mu_+=E_{n+1}(T)-E_n(T)$ and $\mu_-=E_n(T)-E_{n-1}(T)$. The charge gap determines the stability of an $n-$electron state compared to an equal admixture of $(n+1)-$ and $(n-1)-$electron states (the average number of electrons for this mixture is still $n$). Depending on the strength of the on-site electron-electron repulsion $U$, this charge excitation gap can be positive $\Delta^c>0$ or negative $\Delta^c<0$. The pair binding energy $\Delta^c(n,T)$ determines whether a total 2$n$ electrons on pair of isolated clusters prefer to distributed uniformly ($\Delta^c>0$) or give rise to local charge inhomogeneity  ($\Delta^c<0$). For odd $n$ this corresponds also to an effective electron pairing, since clusters are more likely have an even number of electrons. When the charge gap is negative, the system energetically prefers the mixed state with $n+1$ and $n-1$ electron configurations.

The spin gaps $\Delta^h$,  defined in the grand canonical ensemble, can provide essential insights into the response of the system to applied magnetic field $h$. The magnetic susceptibility $\chi_h={{\frac {\partial  S^{z}(\mu)} {\partial h}}}$ is defined as a function of the magnetic field $h$ near the critical chemical potential (say $\mu=\mu_c$) and, in the negative charge gap region ($\Delta^c<0$), we introduce the spin gap $\Delta^h$ as the magnetic field which gives rise to the first magnetic susceptibility peak, {\it i.e.,} the magnetic field required to drive the system into its lowest exited state.

\subsection{Quantum critical points}

The zeros of canonical and the grand canonical gaps as a function of $U$ at $T\to 0$ describe QCPs for continuous and abrupt quantum phase transitions. The change of sign in canonical gaps at the level crossing point away from half filling is crucial for electron (charge) phase separation and pairing instabilities. As temperature approaches zero, the possible sign change in canonical gap $\Delta^c(U_c)=0$ at $U_c$ signifies the existence of a QCP related to first order (dramatic) changes. This QCP describes an abrupt (discontinuous) change in the ground state of a many-body system due to strong quantum fluctuations. The negative charge gap ($\Delta^c<0$) shows the regions of electron instability with a tendency toward phase separation. For one hole off half filling the Mott-Hubbard like insulating ground state for electron-hole pairing with $\Delta^c>0$ is stable at all $U>U_c$, while for $0<U<U_c$, the ground state consists of paired electrons ($\Delta^c<0$). This pairing instability which redistributes electrons signals the formation of spatially inhomogeneous, hole-rich $d^{n-1}$ and hole-poor $d^{n+1}$ regions.

\subsection{Phase separation instabilities}~\label{instabilities}
Many cooperative phenomena invoked in the approximate treatments of the ``large" concentrated systems, are also seen in exact analysis of pairing instabilities in the canonical and grand canonical ensembles of the small clusters in the thermodynamic equilibrium. The negative sign of the canonical charge gap ($\Delta^c<0$) is a signature of phase separation instabilities. However, instead of a full phase separation, the spin gap rigidity ($\Delta^{h}>0$) provokes  local inhomogeneities in electron distribution. The spatial inhomogeneities in the charge distribution imply different electron configurations, close in energy. The quantum mixing of these hole-rich $d^{n-1}$ and hole-poor $d^{n+1}$ clusters for one hole off the half filling case provides a stable spatial inhomogeneous media that allows the pair charge to fluctuate. Thus, the negative sign of charge gap implies phase (charge) separation of the clusters on hole-poor and hole-rich regions.

\subsection{Coherent pairing}
We find that at rather low temperatures the calculated positive gap $\Delta^h>0$ in grand canonical method can have equal amplitude with a negative charge gap $\Delta^c$ derived in the canonical method, {\it i.e.}, with identical gaps, $\Delta=\Delta^h=-\Delta^c$. Such behavior is similar to the existence of a single quasiparticle gap in the conventional BCS state. We identify such an opposite spin (singlet) coupling and electron charge pairing as spin coherent electron pairing. This is an important characteristic of (quantum) phase coherence in the ground state associated with simultaneous coherent pairing of independent charge and spin (Fermion) entities and the corresponding full Bose-Einstein condensation of these ``preformed" bosons with a single energy gap. Such behavior is similar to  coherent pairing with a unique quasiparticle gap in the conventional BCS theory.

\subsection{Incoherent pairing}
However, unlike in the BCS theory, the charge gap differs significantly from the spin gap as temperature increases above the superconducting temperature, $T_c$. It is important to note that the coherent pairing formation through {\sl separate} condensation of bound holes and coupled opposite spins (both boson entities) at different critical temperatures ${T_c}^P$ and ${T_s}^P$ is somewhat different from the Cooper pairs (composite bosons) condensation with a unique critical temperature. The magnitudes of the spin gap and charge gap also differ with increasing temperature here, unlike in the BCS case. This implies the emergence of two gaps and two energy scales or two characteristic critical temperatures leading correspondingly to incoherent pairing of pairs with unpaired (opposite spins) far above $T_c$ seen in HTSCs using STM measurements ~\cite{Yazdani,Hudson,Tanaka,Lee}. Microscopic spatial inhomogeneities and incoherent pairing detected in these experiments are correlated remarkably with predictions of charged pairs without quantum phase coherence above ${T_s}^P$ in nanoscale. The positive change gap $\Delta^c>0$ describes charge redistribution and Mott-Hubbard like transition above ${T_c}^P$. The positive spin gap calculated in the grand canonical approach implies coherent pairing with a homogeneous spatial distribution below ${T_s}^P$. This picture is consistent with some STM measurements in $Bi_2Sr_2CuO_{6+x}$~\cite{Boyer}.

\section{Results}~\label{result}

\subsection{Square pyramids}~\label{sec1}

\begin{figure}
\includegraphics*[width=20pc]{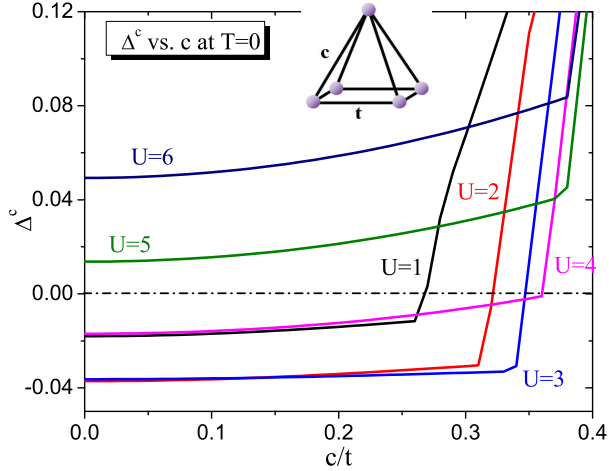}
\caption{A plot of charge energy gap versus intersite coupling $c$. When hoppings between the apical site and the base sites are all the same $t'_{i}=c$, the charge gap $\Delta^c$ is calculated as a function of $c$ at several $U$ values. The region $\Delta^c<0$ has a ground state with coherent pairing $\Delta^c=-\Delta^h$. At relatively small $c$, the gap remains almost constant, which indicates that when the apical site is far from the base, the electronic state on the apical site cannot affect the pairing state on the base. When $c$ is large enough, the dependence of $\Delta^c(c)$ becomes rather steep and beyond the critical value ($c_{crit}$) gap $\Delta^c$ becomes positive. This implies that coupling between the top and base (when apical site gets closer to the plane) can destroy electron pairing at the base. Note that at $U=3$, the system has the largest magnitude of the negative gap covering the widest range of $c$.}
\label{fig:fig2}
\end{figure}

High temperature superconductors have complex layered structures. For example, there are  various layers, bilayers or trilayers that affect the superconductivity attributed to the Cu-O planes acting as {\sl reservoirs} of electrons or holes. Here we attempt to model such complex structures with a simple pyramid having a square base and and an out-of-plane atom. As discussed in our previous publications, a negative charge gap $\Delta^c$ is essential for having coherent pairing states at zero temperature $T=0$. We first consider the simplest case to testify the validity of the model: a symmetric pyramidal cluster with equal hopping terms $t'_i=c$  $(i=1,2,3,4)$ between the apex atom and the all the basal atoms. The charge gap $\Delta^c$ at $T=0$ as a function of coupling parameter $c$ in Fig.~\ref{fig:fig2} is shown at various $U$ values and low temperatures, $T\rightarrow 0$ (in units of $t$) close to optimal doping for one hole of half filling, $N=4$ ($n=0.8$). The energy gap vanishes in a multiparameter space of $U$ and $c$,  where its sign changes define the quantum critical points for electron charge and spin pairing instabilities. In the Fig.~\ref{fig:fig2} plots with $U\leq 4$ provides strong evidence for the existence of a QCP of level crossing instability, associated with phase separation instability. For example, the charge gap at $U=3$ vanishes ($\Delta^c(c_{crit})=0$) at $c_{crit}={t^\prime\over t}\approx 0.35$. This QCP describes an abrupt change in the ground state of a many-body system due to strong quantum fluctuations, {\it i.e.,} a transition from electron pairing into  Mott-Hubbard behavior. The negative charge gap ($\Delta^c<0$) at $0<c<c_{crit}$ corresponds to coherent electron pairing $\Delta^s=-\Delta^h$ in a vertically deformed square pyramid (with $t=1$ in the basal plane), while for $c>c_{crit}$, the Mott-Hubbard insulating ground state for electron-hole pairing with $\Delta^c>0$ is stable.

As shown in Fig.~\ref{fig:fig2}, as the hopping $c$ increases but still small (for example, $c<0.34$ for $U=3$), the value of the charge gap changes very slowly, which corresponds to the case with the top site far from the basal plane. The result indicates that the presence of the top site induces very small perturbations on the electronic state at the planar electronic state when it is far away. However, when $c$ is large enough (for example, $c>0.34$ for $U=3$), there is a dramatic change and the charge gap becomes positive. This illustrates the case with the top site close enough to the base, so the electronic state on the top site greatly affects the electronic state on the basal plane and tends to destroy the electron pairing on the basal as the apical atom gets closer to the plane. Comparing with results of the elementary square geometries~\cite{PRB_08}, we can conclude that the electronic structure of the basal sites contributes significantly to the formation of electron pairs, while the apical site tries to destroy such pairing by ``sharing" holes/electrons with the basal sites. This result is consistent with the conclusion of Slezak et al.~\cite{Slezak} who stated that the ``pair density" is anti-correlated with the height of apical $O$ atom. Given the consistency of this simple result, the pyramid may be a good candidate for studying local gap modulations. The pairing gap has the maximum magnitude and the widest range at $U=3$, so we here consider the gap modulation for only $U=3$.

Another feature of the pairing gap behavior in the pyramidal structure is that when the variations of $c$ are small, the gap varies linearly with a very small slope. It suggests that the apical effects at small $c$ values can be treated by perturbations. This feature resembles the effect of small next nearest neighbor hopping in the square lattice. Actually, the two effects are equivalent when $c$ is small (See Appendix~\ref{s3}). This equivalence indicates that the influence of a far-away top site is like a bridge connecting the next nearest neighbor sites so that it perturbs the electronic state at the plane but does not significantly destroy the paired state. Slezak et al. observed a gap modulation of about $9\%$ in their experiments which is large enough to address the destruction of electron pairs~\cite{Slezak}. Such a big gap modulation is not suitable for a perturbation treatment.

\begin{figure}
\includegraphics*[width=20pc]{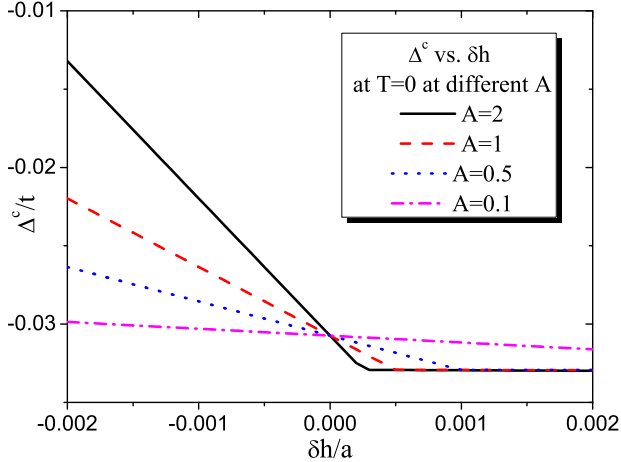}
\caption{The charge gap $\Delta^c$ is plotted as a function of the vertical displacement of the apical site from its equilibrium position at various $A$ values ($A$ is the gradient of hopping $t'$ on the vertical direction). The hopping parameter between apex and base at the equilibrium position is $t'_0=0.34t$. The plot shows two regions with different linear slopes. The apical site should be close enough to the base so that it can provide sufficient sufficient effect on the electronic state on the base. In the figure, this requires $\delta h$ to have a large slope. As $A$ increases, the linear regime with higher becomes smaller. For $A=0.5$, the linear range with higher slope up to about $\delta h=0.001$ yields a modulation amplitude of about $10\%$ of the gap at equilibrium position which means that $A<0.5$ is the suitable region for the simulation.}
\label{fig:fig3}
\end{figure}

\begin{figure}
\includegraphics*[width=20pc]{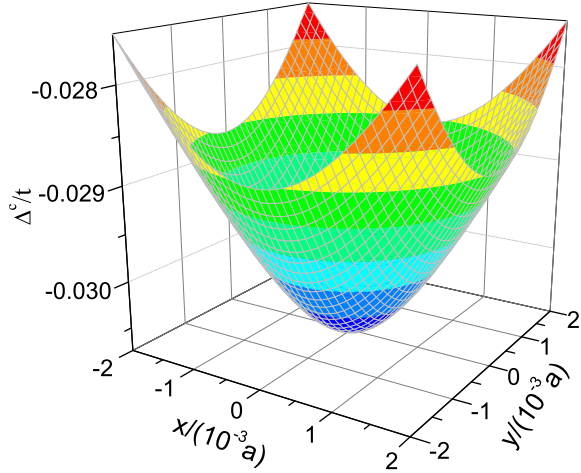}
\caption{Charge gap $\Delta^c$ as a function of the planar $(x,y)$ coordinates of the apical site assuming $z$ (vertical coordinate) is fixed. The gradient of hopping $t'$ on the $x$ and $y$ direction of the basal plane $B$ is set to $0.5$ following the result in Fig.~\ref{fig:fig3} and at the equilibrium position $(0,0)$, which is the center of the basal plane , the hopping between apex and base is set to $t'_0=0.34$. The gap modulation along the horizontal direction is not linear. The amplitude of changes of the gap is about $10\%$ of the gap magnitude at the equilibrium position. The amplitude is close to that in vertical displacements, this suggests that the abilities to interact with basal electronic states caused by vertical and horizontal displacements are almost the same. However, considering the symmetry of lattice, the displacements on the horizontal direction are much more difficult, so apical site mainly oscillates along the vertical direction.}
\label{fig:fig4}
\end{figure}

\subsection{Displacement of apical site}~\label{displacement}

From the analysis of the previous section, we conclude that the best simulation conditions for the charge gap are $U=3t$ and $t'>0.34t$. The next step is to find out the position dependence of the hopping parameters. The hopping parameter between the top and the base is closely related to the distance between the apical site and the basal site, {\it i.e.} $t'_i=t'(r_i)$, where $r_i$ is the distance between the apical site and the basal site $i$. In an ideal crystal, there is an equilibrium position $r_i$ with the hopping parameter $t'_0$ that makes the system most stable. When the layers of atoms are present above the Cu-O planes, the position of the top site can be perturbed with a lattice distortion represented by a very small displacement $\delta r$. The new hopping $t'(r_i+\delta r)$ can be expanded as
\begin{eqnarray}
t'(r_i+\delta r)\approx t'_0+\alpha\cdot\delta r,
\label{eqn:pert}
\end{eqnarray}
where $\alpha$ is the slope of $t'$: $\alpha=\frac{\delta t'}{\delta r}$, so the value of $\alpha$ is strongly dependent on the functional form of $t'(r)$ and the value of $r_i$. Notice, when the variation of the position of the top site is relatively small, the change of the hopping parameter can be well approximated by a linear function. By adjusting the value of $\alpha$ and range of $\delta r$, we can easily model a $9\%$ gap variation.

The variation of the apical site can be the displacement along vertical and/or horizontal directions. A vertical displacement $\delta h$ is equivalent to the case discussed in the Sec.~\ref{sec1} except for converting the dependence of the hopping parameters to that of a spatial displacement. Because $\delta h\approx\frac{r_0}{h}\delta r$,
\begin{eqnarray}
t'(h+\delta h)=t'_0-A\cdot\delta h,
\label{eqn:perth}
\end{eqnarray}
where $A=\alpha\cdot\frac{r_0}{h}=\frac{\delta t'}{\delta h}$. $A$ is the gradient of hopping $t'$ along the vertical ($z$) direction. The pairing gap can be calculated as a function of the vertical displacement based on Eq.~(\ref{eqn:perth}). We choose $t'$ to be $0.34$ and confine the change of the position of apical site $|\delta h|<0.002a$. Fig.~\ref{fig:fig3} plots the charge of gap $\Delta^c$ as a function of $\delta h$ at different $A$ values. In order to simulate the sinusoidal gap behavior, we consider $\Delta^c(\delta h)$ as a linear function of $\delta h$. The plots in Fig.~\ref{fig:fig3} show two regions with different slopes, but only in the region with higher slope, the apical atom is close enough to interact with the basal electronic pairing states. Therefore the useful region for our simulation is only confined to a small region of displacement $\delta h$. As discussed in Appendix~\ref{s5}, we need to have the range of linear function smaller than $|\delta h=0.001|$ in the simulation. According to Fig.~\ref{fig:fig3}, at $A=0.5$, the linear range with higher slope up to (approximately) $\delta h=0.001$ achieves an amplitude about $10\%$ of the equilibrium gap magnitude, which is close to the result in the experiment ($9\%$), so $A<0.5$ is probably the suitable region for $A$ in the simulation.

At fixed $h$ the horizontal distortion provide the relationship between the distance change of apical atom with regard to the atoms in basal plane or corresponding planar $\delta x$ and $\delta y$ displacements
\begin{eqnarray}
\delta r_1&\approx&\frac{a}{2r_0}(-\delta x+\delta y), \nonumber \\
\delta r_2&\approx&\frac{a}{2r_0}(-\delta x-\delta y), \nonumber \\
\delta r_3&\approx&\frac{a}{2r_0}(\delta x-\delta y), \nonumber \\
\delta r_4&\approx&\frac{a}{2r_0}(\delta x+\delta y),
\label{eqn:xy}
\end{eqnarray}
and the corresponding hopping parameters between the apex and the basal sites are reduced to
\begin{eqnarray}
t'_1&=&t'_0-B\cdot(-\delta x+\delta y), \nonumber \\
t'_2&=&t'_0-B\cdot(-\delta x-\delta y), \nonumber \\
t'_3&=&t'_0-B\cdot(\delta x-\delta y), \nonumber \\
t'_4&=&t'_0-B\cdot(\delta x+\delta y),
\label{eqn:pertxy}
\end{eqnarray}
where $B=\alpha\cdot\frac{2r_0}{a}$. $B$ is the gradient of hopping $t'$ on the $x$ and $y$ direction of the basal plane. We confine the change of $x$ and $y$ to a small range $|x|<0.002a$ and $|y|<0.002a$ and $t'$ to be $0.34$ for the same reason. We calculated the charge gap of the system at different horizontal positions $(x,y)$ with $B=0.5$ in Fig.~\ref{fig:fig4}. The gap behaviors with horizontal displacements are not linear and the amplitude of the gap change is also about $10\%$ of the gap magnitude at $(0,0)$ position. By Comparing  Fig.~\ref{fig:fig3} and Fig.~\ref{fig:fig4}, the displacements on vertical and horizontal directions have almost the same effects on the basal electronic states, so the gap change due to the planar displacements appears to be equally important to that caused by vertical distortions. The simulation become rather cumbersome by combining linear vertical and nonlinear horizontal gap behaviors. However, the measurements clearly show the sustained square atomic structure in the $CuO_2$ plane while other layers are strongly distorted by supermodulation~\cite{Kan}. This implies that the planar structure of our model should somehow preserve a square symmetry. Therefore, the horizontal displacements of the apical site must be strongly limited to maintain such a square symmetry. Since the vertical distortions basically do not influence the planar square symmetry, then only vertical $\delta h$ displacements of the apical site are essential in practice for simulation of pairing gap modulation.

\begin{figure}
\includegraphics*[width=20pc]{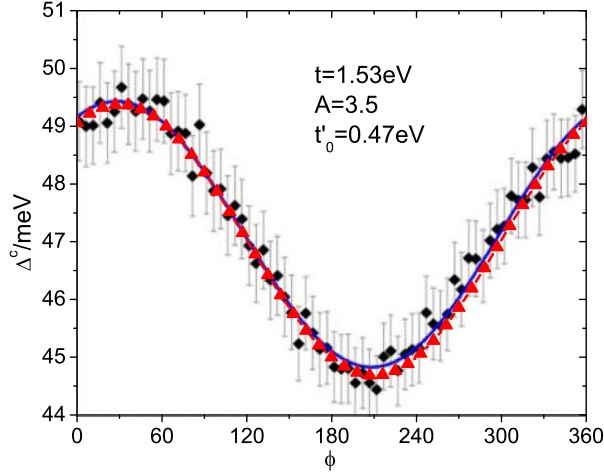}
\caption {Gap modulation due to lattice structure supermodulation where $\phi$ is defined as the phase of lattice structure supermodulation which can be expressed as a sine wave function. The black diamonds with error bars are experimental data with uncertainties from Ref.~\cite{Slezak}. The red triangles are fitted data from our model. The fitting gives the following  parameters for modeling:  $A=0.35$, $t=1.53eV$ and $t'_0=0.47eV$. Note that there is an excellent agreement between experiment and the results from the simple microscopic model.}
\label{fig:fit}
\end{figure}

\subsection{Pairing modulation}

Below we show the results of numerical calculations of pairing gap variation within individual unit cells for various displacements of the apical sites on superconducting pairing. The pairing (sinusoidal variation of gap) modulation has been found in $Bi_2Sr_2CaCu_2O_{8+\delta}$ using scanning tunneling microscopy. The local pyramidal structures affect the superconducting crystalline ``supermodulation" of the energy-gap in bulk $Bi_2Sr_2CaCu_2O_{8+\delta}$. Such supermodulation displacements can be seen for dopant apical oxygen displacement in both the ``a" and ``c" directions and tilt the $CuO_5$ pyramid. The local gap modulations here can merely be related to the dopant density modulation with the same period as the supermodulation. The $CuO_5$ pyramidal coordination of oxygen atoms surrounding each copper atom in $Bi_2Sr_2CaCu_2O_{8+\delta}$ (see Fig. 1 in Ref.~\cite{Slezak}) is equivalent to the square pyramid $CuO_2$ geometry built on the basal $Cu$ squares by a rotation of 90 degrees with respect to the vertical axis). Thus a translation along ``a" direction for $CuO_5$ in $Bi_2Sr_2CaCu_2O_{8+\delta}$ is equivalent to displacements along the diagonal direction in the basal Cu unit cell. The local supermodulation displacements in both the ``a" and ``c" directions are equivalent to vibrational distortions along the diagonal directions in the square pyramid and vertical ``c" direction. Experimental results show explicitly that variations found near the dopant atoms as the primary effect of the interstitial dopant atom which is to displace the apical oxygen so as to diminish the distance ``d" between atoms or distort (slightly tilt) the $CuO_5$ pyramid. We apply our exact results on the square pyramids near the optimal $U=3$ values to fit into the data for the experimentally observed variation of gap with the unit cell distortions.

We study the impact of the apical atom on superconductivity by monitoring the local negative gap as a function of the vertical displacement $\delta h$ of the apical atom from its average position. Experiments have shown that the modulation of the lattice structure can be expressed as a sinusoidal wave function with respect to a phase $phi$ which is defined according to the period of the supermodulation. To simulate the modulation, we introduce the modulation of top site along the vertical direction: $h(\phi)=h_0+\delta h \cdot \sin(\phi+\delta)$, where $h_0$ is the equilibrium position of the apical atom where the hopping between top and base is set to $0.34t$. $\phi$ here is a phase according to the period of the modulation of the top site and $\delta$ is a constant phase shift. We use the absolute value of charge gap as the pairing gap, so we can derive the pairing gap modulation $\Delta(\phi)$ as a function of $\phi$ and by properly setting parameters we can fit our results into experimental data. The fitting with experimental observation of the sinusoidal gap behavior is shown in Fig.~\ref{fig:fit} for optimally doped conditions. The details of the fitting procedure are given in Appendix~\ref{s5}. Our calculations are consistent with results in the Ref.~\cite{Slezak} which shows a $9\%$ modulation of average gap value for optimally doped $Bi_2Sr_2CaCu_2O_{8+\delta}$. The fitting parameters are $t=1.53eV$ and $A=0.35$. $A$, one of the free parameters, is within the range $A<0.5$ so it is consistent with our discussion in the Sec.~\ref{displacement}. The value of $t$, which is the other free parameter, is close to values derived from other theoretical methods~\cite{McMahan, Newns}. This provides additional evidence for the validity of our model.

One interesting question has to do with the intrinsic origin of this pairing gap modulation. There are several possible answers here: the modulation of (1) electron/hole pair strength, (2) the electronic density and (3) the electron/hole pair density. In order to verify the first option, we calculate the change of the chemical potential $\mu$ of the system with the modulation, where $\mu$ is calculated as an average of $\mu_+$ and $\mu_-$ (defined in Sec.~\ref{Method}): $\mu=(\mu_++\mu_-)/2$. The chemical potential represents the average energy needed to break an electron/hole pair, so it shows the average binding energy of the hole/electron pairs. We find that modulation amplitude of $\mu$ is smaller than $0.1\%$ of the average chemical potential. This result indicates that although the pairing gap experiences a strong modulation, the binding energy of the hole/electron pairs is not affected, i.e., the pairing gap modulation is a strong coherent effect not strictly relevant to the modulation of (hole/electron) pairing strength. We have also studied the change of electron density on the apical site with modulation. When the apical site is very far away from basal plane (all $t_i\to 0$), the gap is maximum and there is exactly one electron on the apical atom, i.e., decoupling of the apical atom with the plane restores the two-dimensional character of electron pairing. For our simulated case, when the apical site is spatially close to the plane, we find that the electron density of the top site increases, which supports the experimental observation that the dopant on oxygen can provoke a transfer of holes/electrons on both $CuO_2$ plane and $O$ atoms. However, our calculations show that the amplitude of modulations of the electronic density on apical site is smaller than $1\%$ of the average value. This result indicates that the electronic density on the pairing plane is insignificantly affected by modulation of the out-of-plane structure. Therefore, the gap modulation does not originate from the modulated electronic density. Therefore, we propose the modulation of electron/hole pair density is the source of the pairing gap modulation. This kind of modulation of pair density is somewhat similar to the hole-pair density wave proposed in Ref.~\cite{Chen}.

\section{Summary}~\label{summary}

The effects of disorder in the position of the apical atom outside the CuO$_2$ plane on the superconducting gap have been investigated in $Bi_2Sr_2CaCu_2O_{8+\delta}$ (Bi2212) within the framework of small single band Hubbard clusters. We have shown that exact cluster calculations in appropriate geometries can quantitatively describe the displacements of the atoms in the bulk of $Bi_2Sr_2CaCu_2O_{8+\delta}$ material with its well-known supermodulation behavior. Our calculations at low temperatures can be used to fit and understand experimental data under variation of the position of the apex atom in the vertical ``c" direction or along ``a", ``b" or square diagonal directions. Independent of the character of the quasiperiodic distortion,  we find modulation in the variation of the pairing gap around its central position. The theory provides strong evidence that the supermodulation is correlated with the out-of-plane displacement of the apical site and shows how variations of the inter-atomic distances affect the coupling and maximum of the superconducting energy gap. The current, exact solution-based study offers strong evidence that the inhomogeneity observed in the STM experiments is driven by distortions of the dopant oxygen atoms, located away from the CuO$_2$ plane, whose primary effect is to suppress local pair formation.

The structural changes in our microscopic model in the vicinity of QCP provide the locally enhanced pairing modulation where the nanoscale variation of the pairing strength in the superconducting state have a dramatic effect on the pairing gap. $Bi_2Sr_2CaCu_2O_{8+\delta}$ type materials contain the CuO$_2$ planes and a "reservoir" of carriers in other layers. A charge redistribution may occur between $CuO_2$ plane and nonplanar atoms driven by supermodulation, i.e., out-of-plane displacements of the apical atom. Usually one discusses the charge redistribution by changing composition, such as oxygen stoichiometry, etc. Thus, one of the most direct ways to determine the structural changes in couplings can be done through a detailed analysis of the charge distribution and the valence states of the respective  out-of- and in-plane ions. In particular, the position of the apex oxygen or the corresponding bond length (coupling) is a sensitive probe of these effects. To summarize, this effect may explain structural changes such as the shift of the apical oxygen atom under modulation, and therefore one would expect that there should be a charge density modulation accompanying the formation of a modulated gap in $Bi_2Sr_2CaCu_2O_{8+\delta}$ type materials. However, our results strongly suggest that there should not be significant charge transfer from the reservoir into the superconducting plane. Probably the most direct confirmation of the rigidity of charge density distribution can be obtained for the hole concentration from positron annihilation experiments or X ray absorption spectra~\cite{positron, xray}.

The gap modulation or corresponding quantum oscillations in $Bi$-$2212$ can be attained by tuning pressure or doping due to the direct access they provide to the structural changes that can be traced in the vicinity of QCP. These calculations can also be extended  to finite temperatures, different doping levels or electron concentrations or can be applied to slightly different (octahedron, tetrahedron, etc.) geometries. The obvious next step is to see how the charge gap variations change with temperature and its relationship with the critical temperature T$_c$ of the cuprate superconductors to explore methods of enhancing and maximizing T$_c$. In a nutshell, the spectroscopy assessed by cluster studies of repulsive electrons possesses both energy and symmetry information on the local pairing interaction and consequently contains the essential elements for understanding  the variations of superconducting gap, coherent pairing and modulation in inhomogeneous, large systems. Heavily relying on gained insights from superconductivity effects driven by out-of-plane distortions of the apical atoms, one can understand also the role of local out-of-plane impurities tied to the mechanisms of superconductivity in other layered high T$_c$ superconductors, impurities in graphene intercalated compounds, etc.

\section{Acknowledgments}
We are grateful to A. Bishop, J.C. Davis, K. Fujita, H. She, and J.X. Zhu for useful discussions. The authors acknowledge the computing facilities provided by the Center for Functional Nanomaterials, Brookhaven National Laboratory supported by the U.S. Department of Energy, Office of Basic Energy Sciences, under Contract No.DE-AC02-98CH10886. The work was performed also, in part, at the Center for Integrated Nanotechnologies, a U.S. Department of Energy, Office of Basic Energy Sciences user facility at Los Alamos National Laboratory (Contract DE-AC52-06NA25396) and Sandia National Laboratories (Contract DE-AC04-94AL85000).

\appendix

\section{The relationship between the apical and next nearest couplings}~\label{s3}
The effect on the charge gap of a small apical hopping is similar to that of a small next nearest hopping. Here are a little simple investigates of the relation between these two small extra hopping based on the first order perturbation. For the simplest case, assume all the apical hoppings are the same $t'$, and $t_n$ is used to denote next nearest hopping. Both $t'$ and $t_n$ are small relative to $t$.

Assume that the ground state of a four-site cluster without next nearest neighbor is $|\psi\rangle$ and the ground state of an isolated apical atom is $|g\rangle$, so the first order correction due to the next nearest hopping is
\begin{eqnarray}
\Delta E_n=\langle\psi|t_n\sum_{nnn}c^+_ic_j|\psi\rangle
\label{eq:nnn}
\end{eqnarray}
where $\sum_{nnn}$ means summation over all the next nearest neighbors. While, the first order perturbation due to the apical hopping is
\begin{eqnarray}
\Delta E'=\langle g\psi|t'\sum_{nnn}c^+_i(c\sum_{m}|m\rangle \langle m|c^+)c_j|g\psi\rangle
\label{eq:apc}
\end{eqnarray}
where $|m\rangle$ is the $m^{th}$ state of the top atom. Comparing Eq.~\ref{eq:apc} and Eq.~\ref{eq:nnn}, $c\sum_{m}|m\rangle \langle m|c^+$ multiplication is the only difference. If $c\sum_{m}|m\rangle \langle m|c^+ = \mathbf{1}$, the two equation are equivalent. Usually, $|m\rangle$ can be state with no electron $|0\rangle$, one up spin $|\uparrow\rangle$, one down spin $|\downarrow\rangle$, or two spins $|\uparrow\downarrow\rangle$ and these four states build up a complete basis, so $\sum_{m}|m\rangle \langle m|=\mathbf{1}$. However, $c|m\rangle$ only include three terms $|0\rangle$, $|\uparrow\rangle$, and $|\downarrow\rangle$, so
\begin{eqnarray}
\Delta E'=t'\sum_{nnn}(\langle g\psi|c^+_ic_j|g\psi\rangle-\langle g\psi|\uparrow\downarrow\rangle\langle\uparrow\downarrow|g\psi\rangle)
\end{eqnarray}
Therefore, as long as the ground state of apical atom is not a two spin state, the effect on the ground state energy within the first order correction due to the apical atom is same as the next nearest hopping. It is hard to determine all the cases that the condition is satisfied, but some specific cases can be empirically found: ($1$) when hopping between the apex and the plane is small, the holes have very small probability to occupy the apical atom at the same time; ($2$) When the electron interaction is strong enough, the double occupancy is prohibited, but this case also causes the spin frustration because of the non-bipartite structure.

\section{Modulation fit with experiment}~\label{s5}

The experiment (see ~\cite{Slezak}) shows a cosinusoidal superconducting gap behavior. The average gap value at optimal doping is about $47$meV and the (peak-to-peak) amplitude of gap modulation is about $9-10\%$ of the average gap value. The sinusoidal behavior is caused by the super modulation in the structure of the BSCCO system. The detail of the modulation structure can be found in Ref.~\cite{Kan}. From these papers, The $CuO_2$ layer does not change too much from the original square structure, but the $SrO$ and $BiO$ layers are strongly distorted. Therefore, the modulation is mainly coming from the changes of $SrO$ and $BiO$ layers and the modulation of the $Bi$ and $Sr$ atoms are found to be combination of sinusoidal waves.

In our model, the effects of the $SrO$ and $BiO$ layers are included in the apical site of the pyramid Hubbard model. The distortion of the two layers can be simulated by the displacement of the apical site along vertical and horizontal direction. As mentioned in Sec.~\ref{displacement}, only the vertical displacement is essential in our model due to the requirement of a strong square symmetry, so the horizontal distortions can be ignored. The equilibrium hopping term $t_0$ between top and base is set to $0.34t$ at which the pairing gap is about $0.30t$. In the experiment the average gap value is $0.47eV$, so $t=1.53eV$. Experiments~\cite{Slezak,Kan,Bianconi} shows that the vertical amplitudes of the modulation layers are about $0.01c-0.02c$ where $c$ is the vertical lattice constant. The amplitude of vertical displacement of the apical site in our model should be much smaller than experimental results because the simplified apical site averages in matter of fact describe displacements of several (coupled) layers in real material. In the paper, we set the amplitude to $\delta h_max=0.001$. The fitting with experimental results also provides $\alpha=3.5$ and $\delta=\frac{\pi}{6}$ by the amplitude and the phase shift of gap modulation.

\end{document}